# A COMPARATIVE STUDY OF PERFORMANCE OF FPGA BASED MEL FILTER BANK & BARK FILTER BANK

Debalina Ghosh[1], Depanwita Sarkar Debnath[1], Saikat Bose[1]

[1]Department of Microelectronics & VLSI Design, Techno India, SaltLake, Kolkata, India

debolina1512@gmail.com
depanwita_sarkar@yahoo.co.in , troykol@yahoo.co.in

## ABSTRACT

*The sensitivity of human ear is dependent on frequency which is nonlinearly resolved across the audio spectrum .Now to improve the recognition performance in a similar non linear approach requires a front-end design, suggested by empirical evidences. A popular alternative to linear prediction based analysis is therefore filter bank analysis since this provides a much more straightforward route to obtain the desired non-linear frequency resolution. MEL filter bank and BARK filter bank are two popular filter bank analysis techniques. This paper presents FPGA based implementation of MEL filter bank and BARK filter bank with different bandwidths and different signal spectrum ranges. The designs have been implemented using VHDL, simulated and verified using Xilinx 11.1.For each filter bank, the basic building block is implemented in Spartan 3E. A comparative study among these two mentioned filter banks is also done in this paper.*
.

## 1. INTRODUCTION

Psychophysical studies have shown that human perception of the frequency content of sounds, either for pure tones or for speech signals, does not follow a linear scale. The idea of defining subjective pitch of pure tones has been derived from this research. How high the voice is i.e., its pitch, is a psychological dimension that represents how high and low sounds are. Pitch is greatly, but not exclusively, dependent on frequency; the intensity of sounds also affects the perception of pitch. The musical scale of pitch is one measure that represents the psychological quantity of pitch, but octaves and doubling the pitch are not the same sensation. In order to construct a psychological measure for different quantification of the pitch sensation we need to carry out pitch measurement experiments based on the magnitude estimation method like those of some measurements for loudness for determining what is heard is double the pitch and what is heard as half the pitch of a standard sound. One psychological measure of pitch created in this manner is the Mel scale [1]. The Mel scale, named by Stevens, Volkman and Newman in 1937 [2] is a perceptual scale of pitches judged by listeners to be equal in distance from one another. An assigned perceptual pitch of 1000 Mels to a 1000 Hz tone, 40 dB above the listener's threshold value can defines the reference point between this MEL scale and normal frequency measurement. Listeners can produce equal pitch increments over larger and larger intervals approximately above about 500 Hz [2].For that reason two octaves on the Mel scale can be comprised with the four octaves on the hertz scale above 500 Hz[18]. The name Mel comes from the word melody to indicate that the scale is based on pitch comparisons [2].The Mel scale (in KHz) can be approximated by following formula [3].

Mel (f) = 2595 $\log_{10}$(1+f/700)             (1)

The Mel-filter bank is designed to simulate band pass filtering occurring in auditory system such that it is approximately linear up to 1 kHz and in actual frequency domain is logarithmic at higher frequencies [Picone, 1993]. Such a model allows a constant bandwidth and constant spacing on the Mel-frequency scale and exploits the fact that the speech signal is stationary for short periods of time. It is modeled by constructing the required number of triangular band-pass filters with 50% overlap. Triangular band-pass filters are generated with Mel-frequencies to be the centers of the triangles .Figure 1 shows the Mel filter bank.

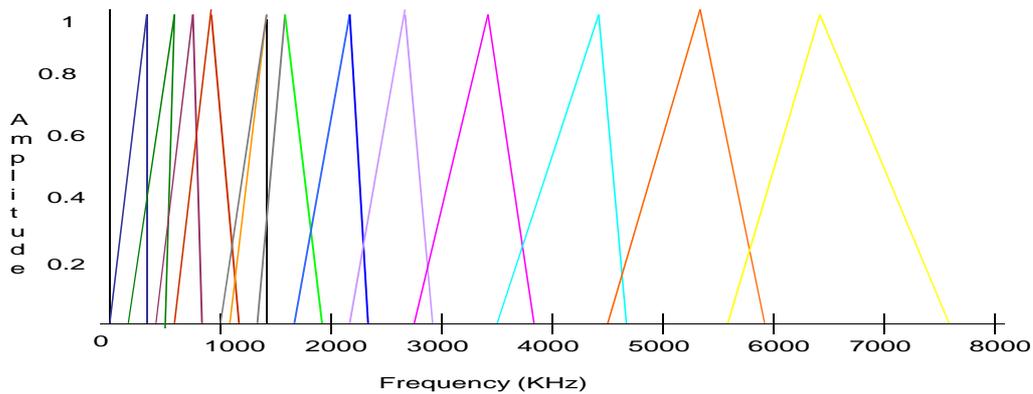

Figure 1.  Mel Filter Bank

The Bark scale provides an alternative perceptually motivated scale to the Mel scale. The basilar membrane (BM) which is an important part of the inner ear performs the spectral analysis followed by speech intelligibility perception in humans. Each point on the BM can be considered as a band pass filter having a bandwidth equal to one critical bandwidth or one Bark [5]. The bandwidth of several auditory filters were empirically observed and used to formulate the Bark scale. An approximate expression (2) for the Bark scale frequency warping, due to Schroeder [2], is used in these experiments

Bark (f) = { 13 atan (0.76f /1000) +3.5 atan ($f^2$/ $(7500)^2$ ) }             (2)

The Bark filter bank is designed to simulate band pass filtering occurring in auditory system such that below 500 Hz the Bark scale becomes more and more linear. Fig 2 shows the Bark filter bank.

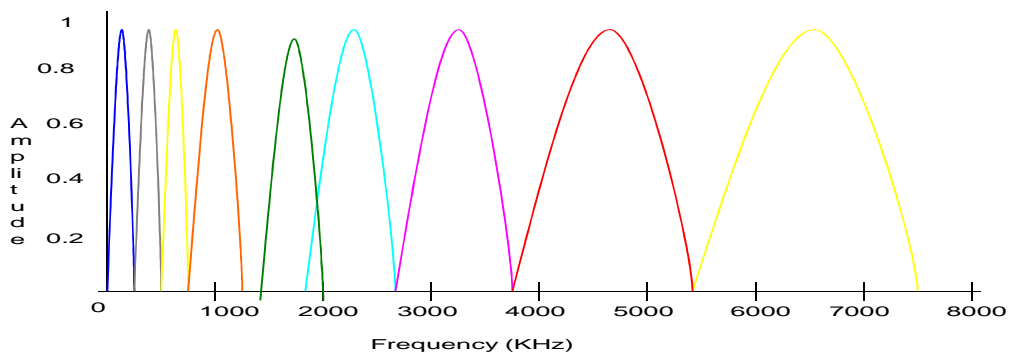

Figure 2 .Bark Filter bank

In this paper we represent Mel filter bank and Bark filter bank using Thirteen Band pass filters. The designs are synthesized in Xilinx VIRTEX 4 device successfully. The synthesis performance results of the designs and the device utilization will also be compared with each other. The next section describes the architectural design of the MEL and Bark filter bank structure. Section 3, shows implementation and design results. Finally, conclusions are exposed in section 4.

## 2. PROPOSED METHODOLOGY

MEL filter bank and Bark filter bank are designed using band pass FIR filter with Bartlett and Hamming window accordingly. In these designs Xilinx Logicore FIR compilers has been used. This design is implemented using Xilinx ISE11.1 and synthesized in FPGA through VHDL.

### 2.1. Architecture of Mel filter bank

The design flow of Mel filter bank is shown in figure 3. Thirteen Fir filters, designed with Bartlett window, are used, having different low pass and high pass frequencies. The design as shown in figure 3 is verified using three different set of frequency filter bank to make a comparative study among all. In Table 1, the lower and upper cut of frequencies of band pass filter is assigned. Table 2 and table 3 shows lower and upper cut off frequencies with different band width ,and different signal range .After that, the output of each filter is passed through full adder to get the output coefficients of MEL filter bank .

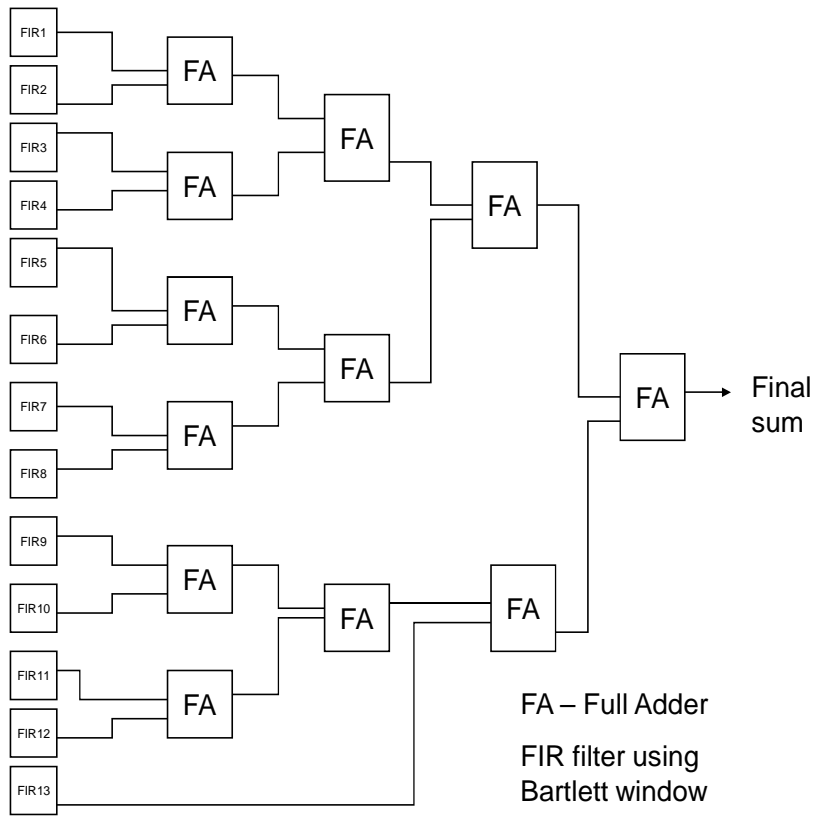

Figure 3. Architecture of Mel filter bank

**Table 1.** Higher and lower cut of frequencies of band pass FIR filter for Mel bank Design architecture1

| No Of Filter | Lower Cut off Frequency (KHz) | Upper Cut off Frequency (KHz) | Bandwidth (KHz) |
|---|---|---|---|
| 1 | 50 | 250 | 200 |
| 2 | 250 | 450 | 200 |
| 3 | 450 | 650 | 200 |
| 4 | 650 | 850 | 200 |
| 5 | 850 | 1062 | 212 |
| 6 | 1058 | 1358 | 300 |
| 7 | 1350 | 1750 | 400 |
| 8 | 1742 | 2262 | 520 |
| 9 | 2256 | 2956 | 700 |
| 10 | 2948 | 3758 | 810 |
| 11 | 3750 | 4700 | 950 |
| 12 | 4692 | 5962 | 1270 |
| 13 | 5960 | 7625 | 1675 |

**Table 2.** Higher and lower cut of frequencies of band pass FIR filter for Mel bank Design architecture2

| No Of Filter | Lower Cut off Frequency (KHz) | Upper Cut off Frequency (KHz) | Bandwidth (KHz) |
|---|---|---|---|
| 1 | 150 | 250 | 100 |
| 2 | 250 | 350 | 100 |
| 3 | 350 | 450 | 100 |
| 4 | 450 | 550 | 100 |
| 5 | 550 | 670 | 120 |
| 6 | 665 | 825 | 160 |
| 7 | 820 | 1060 | 240 |
| 8 | 1055 | 1415 | 360 |
| 9 | 1410 | 1910 | 500 |
| 10 | 1906 | 2606 | 700 |
| 11 | 2600 | 3550 | 950 |
| 12 | 3545 | 3845 | 1250 |
| 13 | 3840 | 5490 | 1650 |

**Table 3.** Higher and lower cut of frequencies of band pass FIR filter for Mel bank Design architecture3

| No Of Filter | Lower Cut off Frequency (KHz) | Upper Cut off Frequency (KHz) | Bandwidth (KHz) |
|---|---|---|---|
| 1 | 10 | 60 | 50 |
| 2 | 60 | 110 | 50 |
| 3 | 110 | 160 | 50 |
| 4 | 160 | 210 | 50 |

| 5 | 210 | 360 | 150 |
| 6 | 340 | 690 | 350 |
| 7 | 670 | 1320 | 650 |
| 8 | 1310 | 2360 | 1050 |
| 9 | 2300 | 3850 | 1550 |
| 10 | 3840 | 5990 | 2150 |
| 11 | 5980 | 6830 | 2850 |
| 12 | 6810 | 7460 | 3650 |
| 13 | 7440 | 11990 | 4550 |

## 2.2. Architecture of BARK filter bank

The design flow of Bark filter bank is shown in figure 4. Thirteen Fir filters, designed with hamming window, are used, having different low pass and band pass frequencies. The design as shown in figure 4 is verified using three different set of frequency filter bank to make a comparative study among all. In Table 4, the lower and upper cut of frequencies of band pass filter is assigned. Table 5 and table 6 shows lower and upper cut off frequencies with different band width, and different signal range after that, the output of each filter is passed through full adder to get the output coefficients of Bark filter bank.

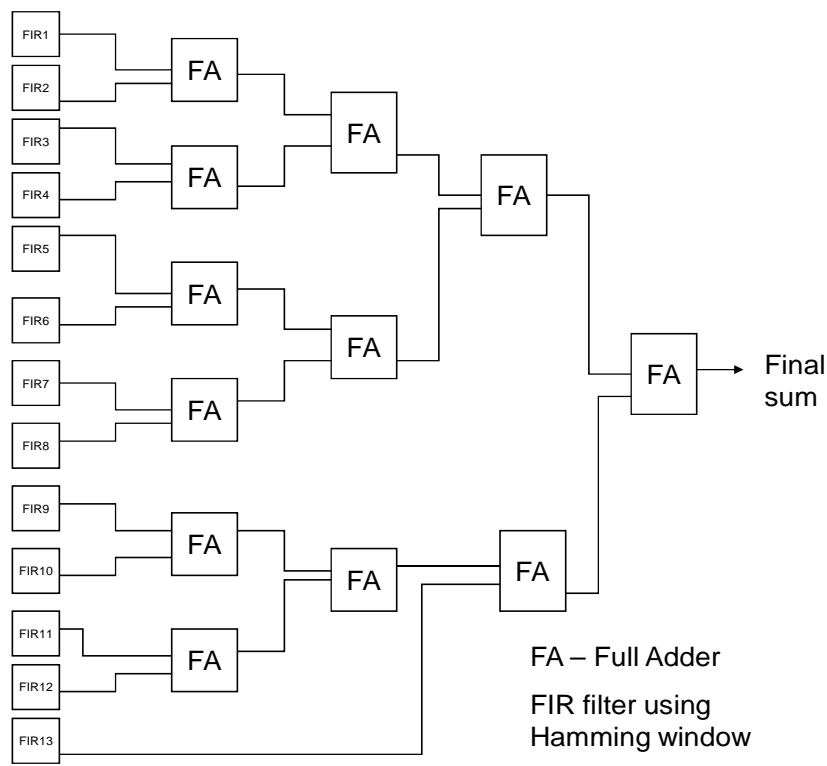

Figure 4 .Architecture Of BARK filter bank

**Table 4.** Higher and lower cut of frequencies of band pass FIR filter for BARK bank Design architecture1

| No Of Filter | Lower Cut off Frequency (KHz) | Upper Cut off Frequency (KHz) | Bandwidth (KHz) |
|---|---|---|---|
| 1 | 50 | 200 | 150 |
| 2 | 200 | 350 | 150 |
| 3 | 350 | 500 | 150 |
| 4 | 500 | 650 | 150 |
| 5 | 650 | 900 | 250 |
| 6 | 900 | 1300 | 400 |
| 7 | 1300 | 1900 | 600 |
| 8 | 1900 | 2750 | 850 |
| 9 | 2750 | 3900 | 1150 |
| 10 | 3900 | 5400 | 1600 |
| 11 | 5400 | 7400 | 2000 |
| 12 | 7400 | 9900 | 2500 |
| 13 | 9900 | 12900 | 3000 |

**Table 5.** Higher and lower cut of frequencies of band pass FIR filter for BARK bank Design architecture2

| No Of Filter | Lower Cut off Frequency (KHz) | Upper Cut off Frequency (KHz) | Bandwidth (KHz) |
|---|---|---|---|
| 1 | 150 | 250 | 100 |
| 2 | 250 | 350 | 100 |
| 3 | 350 | 450 | 100 |
| 4 | 450 | 550 | 100 |
| 5 | 550 | 700 | 150 |
| 6 | 700 | 900 | 200 |
| 7 | 900 | 1200 | 300 |
| 8 | 1200 | 1650 | 450 |
| 9 | 1650 | 2300 | 650 |
| 10 | 2300 | 3100 | 800 |
| 11 | 3100 | 4200 | 1100 |
| 12 | 4200 | 5650 | 1450 |
| 13 | 5650 | 7500 | 1850 |

**Table 6.** Higher and lower cut of frequencies of band pass FIR filter for BARK bank Design architecture3

| No Of Filter | Lower Cut off Frequency (KHz) | Upper Cut off Frequency (KHz) | Bandwidth (KHz) |
|---|---|---|---|
| 1 | 10 | 60 | 50 |
| 2 | 60 | 110 | 50 |
| 3 | 110 | 160 | 50 |
| 4 | 160 | 210 | 50 |

| 5 | 210 | 360 | 150 |
| 6 | 360 | 710 | 350 |
| 7 | 710 | 1360 | 650 |
| 8 | 1360 | 2410 | 1050 |
| 9 | 2410 | 3960 | 1550 |
| 10 | 3960 | 6110 | 2150 |
| 11 | 6110 | 8960 | 2850 |
| 12 | 8960 | 12610 | 3650 |
| 13 | 12610 | 16010 | 4000 |

## 2.3. Design Methodology

Best solution of each design situation can be achieved through Design Methodology. It gives the approaches of a set of conception which, later is used to realize the processing of product. A thoughtful design process can be successfully stretched into actual process design and fabrication for the need of early design and composite conceptual decisions. Modern Technology industry has developed several approaches of Design Methodology. Top down design and bottom up design are two important design methodologies. In this paper Mel filter bank and Bark filter banks are designed using Xilinx Logicore in association with MATLAB. Figure 5 shows the design flow of Filter bank.

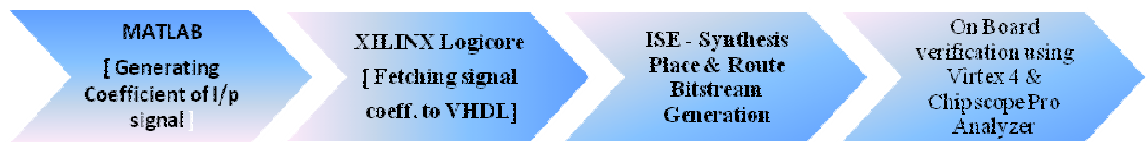

Figure 5. Design flow of Mel filter & Bark filter bank generation

Coefficients file of Hamming window and Bartlett window [.coe file] are generated for band pass FIR filters having different low pass and high pass frequencies .Then that is instantiated into Logicore FIR block which is illustrated in the following Figure 6.

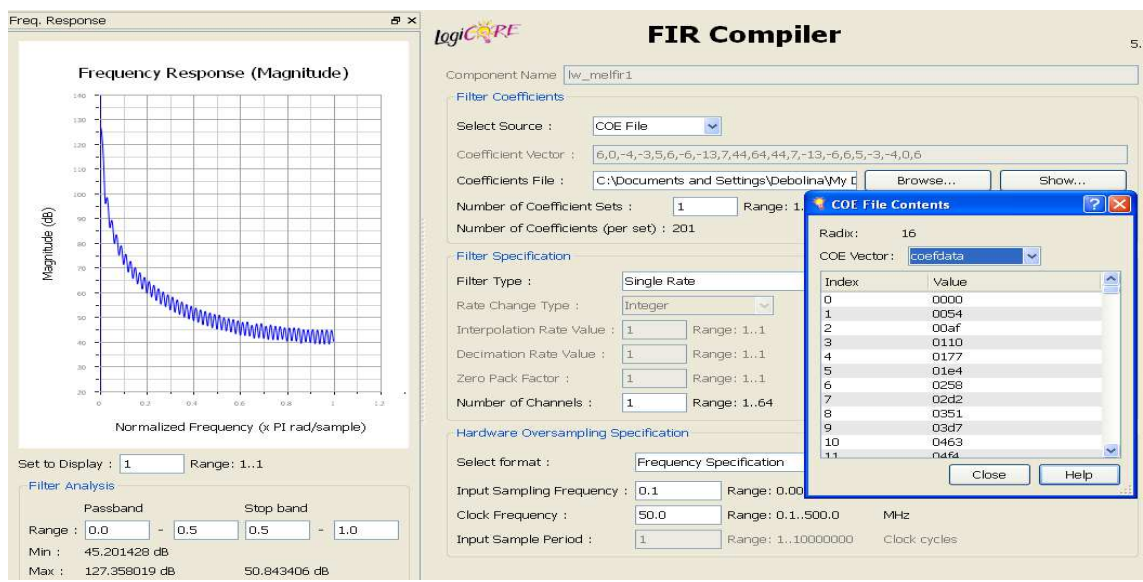

Figure 6. Xilinx Logicore FIR Filter model for lower bandwidth FIR filter

Here Coefficient Vector file is used to specify the filter coefficients. As per the user requirements, FIR compiler itself used to appropriately quantize the non –integer real numbers of filter coefficients with coefficients vectors, specified in the .coe file. Coefficients File is actually the Coefficient file name, which is the file of filter coefficients. A .coe extension is included in the file format which is described in the "Filter coefficient Data"section [17].User can select the file browsing the dialogue box. The sample frequency for one channel is specified by Input Sampling Frequency Field in an integer or in real value. The clock frequency, various filter parameters like Interpolation Rate, number of channels set the upper limit. Clock frequency field can be an integer or real value. Architecture choices are influenced by the limitation of sample frequency, interpolation rate and number of channels .Hence in final implementation, specified clock rate may not be achieved due to these constraints. The value of "Input/output Sample Period" [17] can be assigned depending upon design requirement. It is specified by the integer number of clock cycles between the times division multiplexed input samples and output samples when the multiple channels or fixed fraction decimation filters have been specified respectively. Available clock cycles can be more efficiently used by specifying the output sample period. Set to Display selects which of multiple Frequency Response Graph. In Pass band range two fields are available to specify it, the left-most being the minimum value and the right-most the maximum value. The values are specified in normalized to pi radians per second as on the graph x-axis For the specified range the pass band maximum, minimum and ripple values are calculated and displayed (in dB).In Stop band range two fields are available to specify the stop band range, the left-most being the minimum value and the right-most the maximum value. Here also the values have the same unit as on the graph x-axis i.e. normalized to pi radians per second. Depending upon the design, the stop band maximum value is calculated and displayed in db for the specified range [17]. After that, LogiCore HDL template files are used in ISE to design the ultimate program .Next Program is simulated through Xilinx Isim 11.1 from where we can get the clear view of output coefficients of MEL bank Fir filter and Bark bank Filter .In the last stage both programmes are downloaded to FPGA kit for real time implementation through ChipScope Pro logic Analyzer tool.

## 3. RESULTS, DISCUSSION AND DEVICE UTILIZATION SUMMARY

### 3.1. Results and Discussion

The designs of Mel filter bank and Bark filter bank have been simulated in Xilinx Isim 11.1. Here DDS (Direct digital synthesizer) is used to give input as sine wave to both of these two filter banks. Figure 7, figure 8 and figure 9 shows the output waveform of Mel filter banks designed for different spectral ranges. The outputs are processed during active low reset and high clock enable signal. Internal computed data are get when clock signals goes high to low , that is assigned as stop_the_clock ="FALSE" .The final output is get when stop_the_clock signal is "TRUE". Whole design is simulated for 1200 us. Here the clock frequency is 50 MHz.

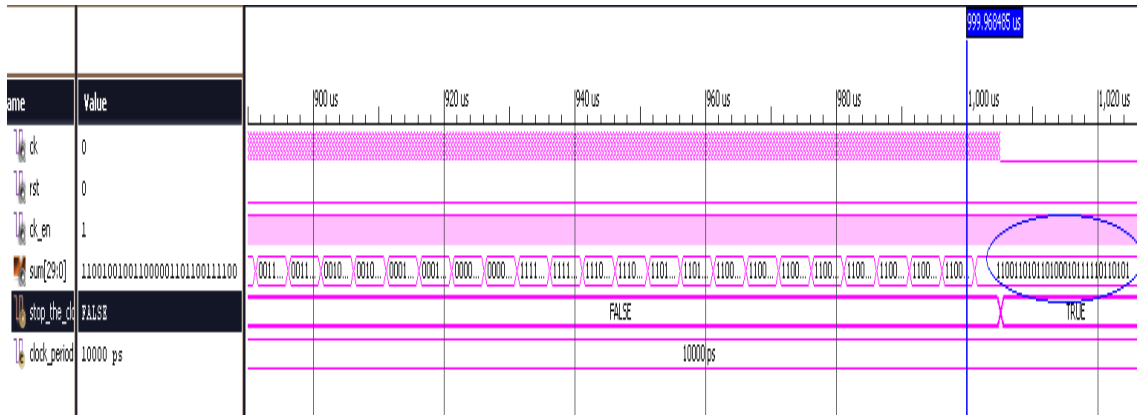

Figure 7 .Simulated output waveform of Mel filter bank1 with spectral range from 50 KHz to 7625 KHz .

.

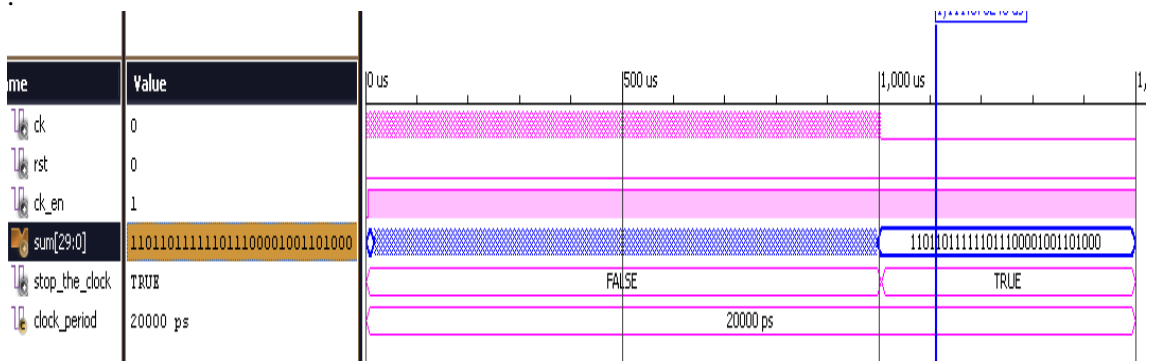

Figure 8. Simulated output waveform of Mel filter bank2 with spectral range from 150 KHz to 5490 KHz

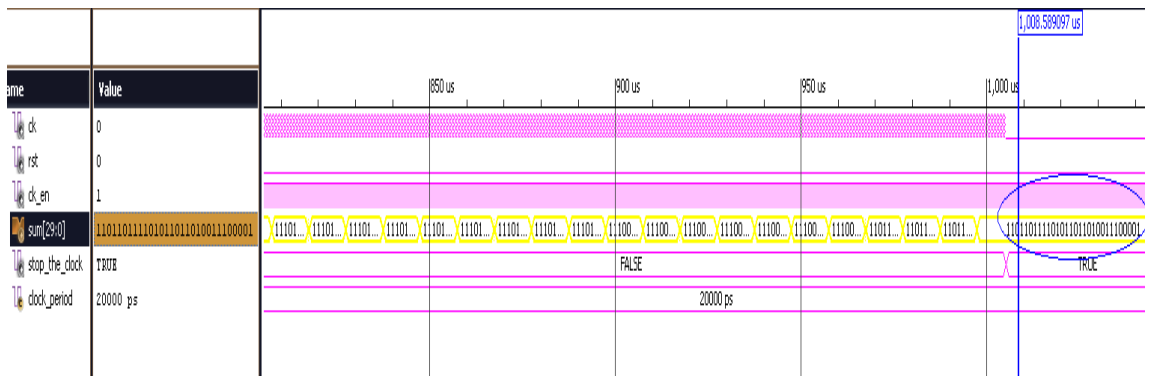

Figure 9. Simulated output waveform of Mel filter bank3 with spectral range from 10 KHz to 11990 KHz

Simulated test bench waveform of Bark filter bank with different spectral ranges are shown in figure 10, figure 11 and figure12. The condition of getting the output waveform of Bark filter bank is same as prior discussed output waveform of Mel filter bank.

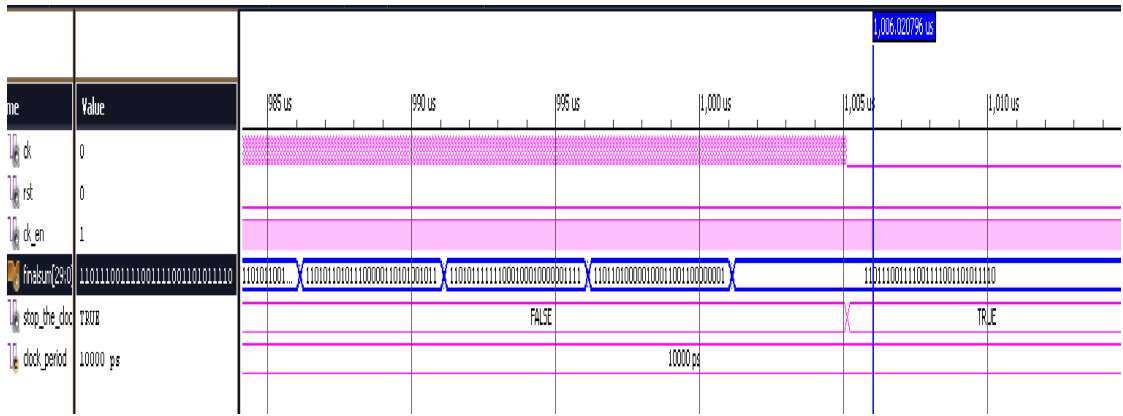

Figure 10. Simulated output waveform of Bark filter bank1 with spectral range from 50 KHz to12900 KHz

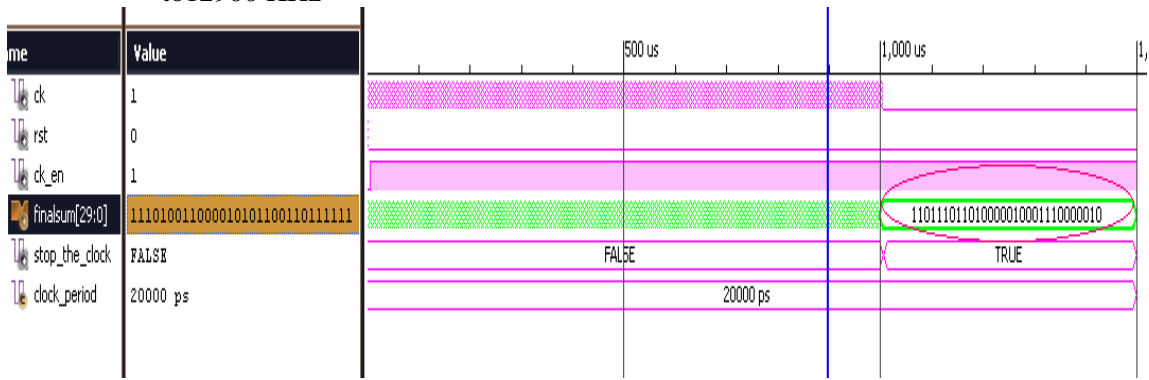

Figure 11 . Simulated output waveform of Mel filter bank2 with spectral range from 150 KHz to 7500 KHz

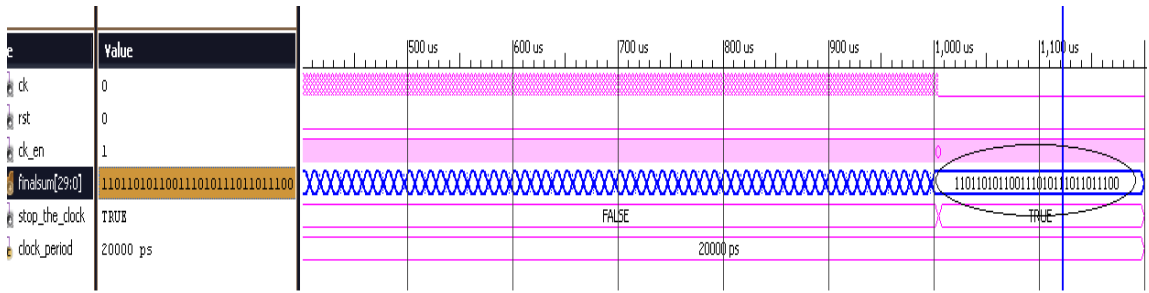

Figure 12 . Simulated output waveform of Mel filter bank2 with spectral range from 10 KHz to 16010 KHz

From the above test bench waveforms of Mel filter banks and Bark filter banks, it is realizable that same output is obtained at the final stage when spectral ranges are quite high but for small spectral ranges the final output differ from each other. Also during the simulation processing, the intermediate computed data of Bark filter and Mel filter are different from each other. Both shows non linear frequency transformation but in computational performance Bark filter is more efficient than Mel filter. Fig 13 shows the comparison of coefficients of Bark and Mel filter bank. From the given figure it is computed that bark filter (65%) higher accuracy than Mel filter (63.37%) and the proposed filter bank (63.92%) referred in [13].

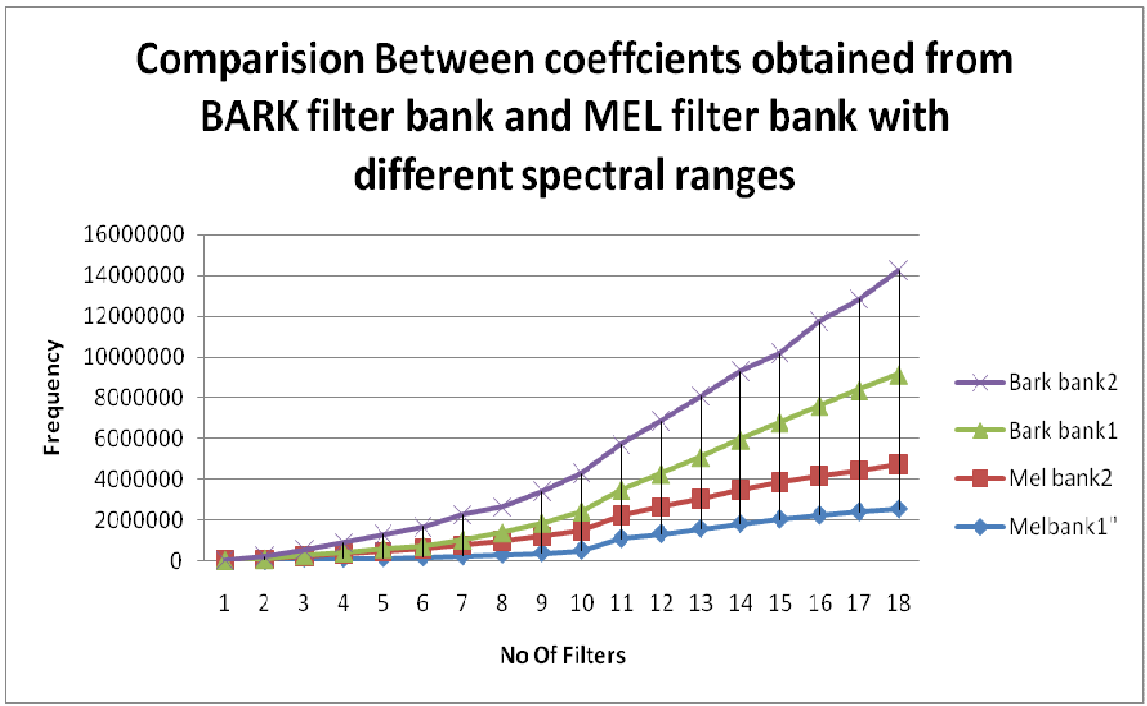

Figure 13. Comparison between coefficients obtained from Bark and Mel filter bank

### 3.2. FPGA Device Utilization Summary

Device Utilization summary is a report which allows designer to view the information like targeted device, the number of errors and warning, device utilization & design goal. We have implemented our design in FPGA family Virtex 4(4vfx12ff668speed grade -12).Also the RTL schematic of Mel filter bank and Bark filter bank are shown. RTL View represents the circuit design graphically, which is an acronym of Register Transfer Level .It is an approach of prior stages of synthesis process when technology mapping is not yet completed. It is generated by synthesis tool, like Xilinx Synthesis Technology produces it as .ngr file. Through the successful RTL design, original HDL code can be nearly achieved. In the RTL view, the design is represented in terms of macro blocks, such as adders, multipliers, and registers. Standard combinatorial logic is mapped onto logic gates, such as AND, NAND, and OR, which are generated when the design become correct in simulation and synthesis level [15] . Individual Fir filter are also implemented in FPGA family virtex 4 and output are checked through FPGA ChipScope pro. All internal signals of the design are monitored through customizable logic analyzer core, named as ILA, acronym of ChipScope Pro Integrated Logic Analyzer. Features of Modern Logic analyzers like Boolean trigger equations, trigger sequences, and storage qualification are also included in ILA which bolsters its advancement in logic analyzing. Because the ILA core is synchronous to the design being monitored, all design clock constraints that are applied to design are also applied to the components inside the ILA core. It has many important features like – It communicates between ChipScope Pro Analyzer software and capture cores through the ChipScope Pro ICON core. It gives designer the flexibility to select trigger width, data width and data depth according to design's requirement. A single trigger condition can be achieved combining its multiple trigger ports. It is also capable to store samples depending on certain condition [16] .This is shown in figure 14 .Figure 15 shows the Zoomed output waveform of Real time realization of Individual Fir filter of Filter bank in ChipScope Pro Analyzer. Figure 16 & figure 17 shows RTL schematic &device utilization summary of Mel filter bank. Figure 18 & figure 19 shows the RTL schematic &device utilization summary of Bark filter bank.

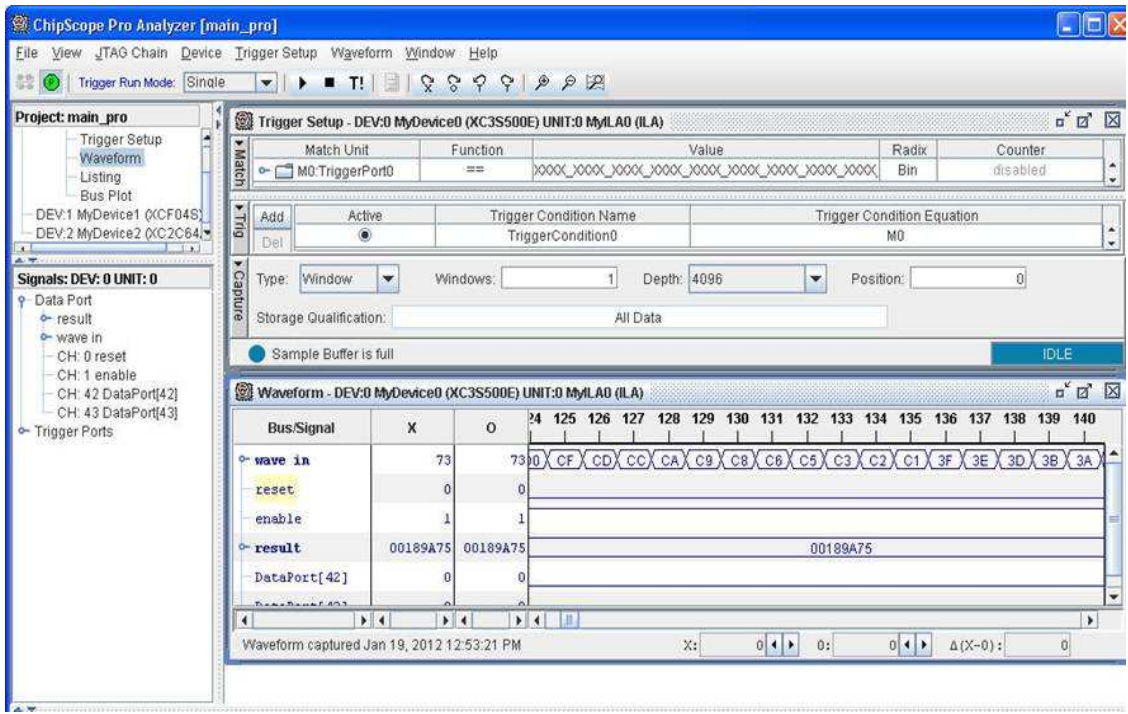

Figure 14. Real time realization of Individual Fir filter of Filter bank in ChipScope Pro Analyzer

From Figure 14 it is observed that when we specify "Trigger run mode", single and hit the immediate trigger option we get the corresponding output to Input. The outputs are also dependent parameter of Reset and Enable. The output is indicated here as "result".

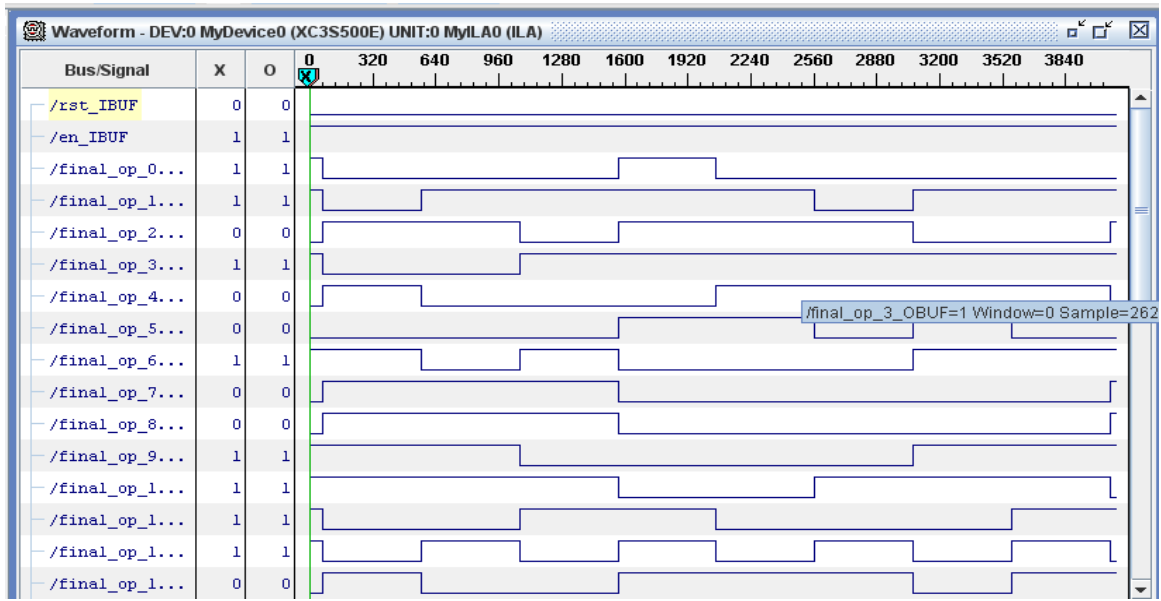

Figure 15. Zoomed output waveform of Real time realization of Individual Fir filter of Filter bank in ChipScope Pro Analyzer

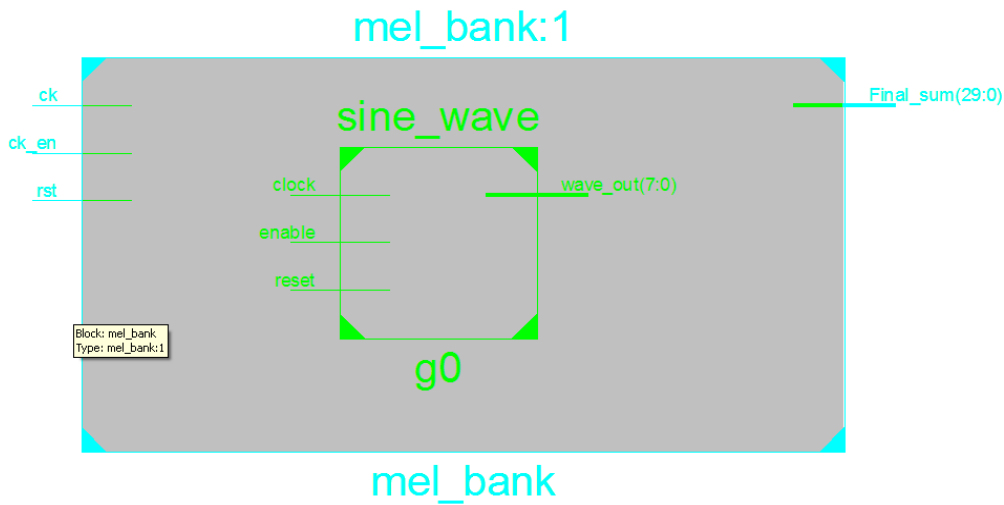

Figure 16 .RTL schematic of Mel filter bank

In Figure 16, it is observable that sine wave block is instantiated in the "mel_bank" block, which is act as input signal spectrum .Clock, Clock_en and reset are global and we get the output of 30 bits corresponding to input of 8 bits. Figure 18 can also be illustrated as figure 16.

| bank_bark Project Status (01/26/2012 - 03:22:51) | | | |
|---|---|---|---|
| Project File: | mel.xise | Parser Errors: | No Errors |
| Module Name: | nwmel_bank | Implementation State: | Synthesized |
| Target Device: | xc4vfx12-10ff668 | • Errors: | No Errors |
| Product Version: | ISE 12.3 | • Warnings: | 51 Warnings (25 new) |
| Design Goal: | Balanced | • Routing Results: | |
| Design Strategy: | Xilinx Default (unlocked) | • Timing Constraints: | |
| Environment: | System Settings | • Final Timing Score: | |

| Device Utilization Summary (estimated values) | | | |
|---|---|---|---|
| Logic Utilization | Used | Available | Utilization |
| Number of Slices | 1134 | 5472 | 20% |
| Number of Slice Flip Flops | 1906 | 10944 | 17% |
| Number of 4 input LUTs | 1833 | 10944 | 16% |
| Number of bonded IOBs | 33 | 320 | 10% |
| Number of FIFO16/RAMB16s | 26 | 36 | 72% |
| Number of GCLKs | 1 | 32 | 3% |
| Number of DSP48s | 13 | 32 | 40% |

Figure 17. Device utilization summary of Mel filter bank

Device utilization of Mel filter bank is illustrated in Figure 17 .The project status block describes whether the design is synthesized ,having different errors or not. Here in our design warnings are notified due to use of Logicore block which can only be able to instantiate structural models. Here design goal is assigned as "balanced" because in "Device utilization summary block (estimated value)" all logic utilizations are under mapped. The same explanation is also valid to figure 19.

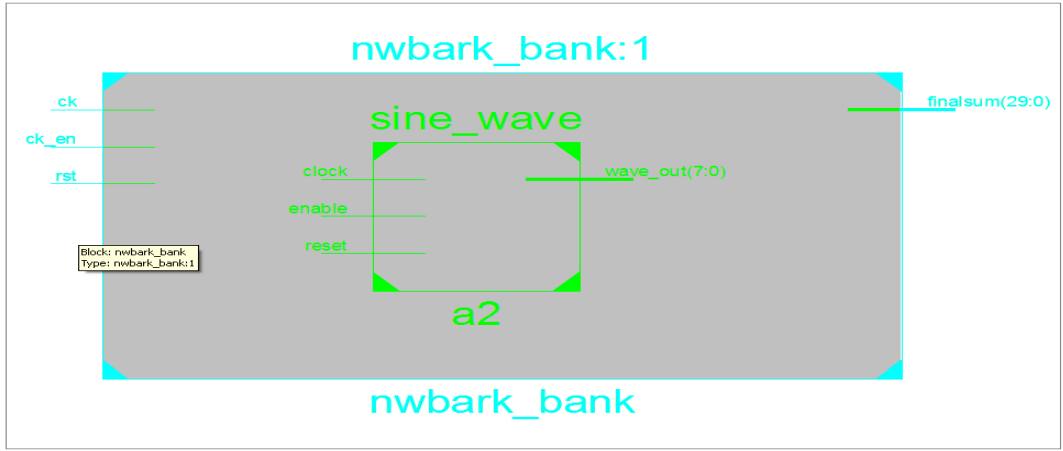

Figure 18 .RTL schematic of Bark filter bank.

| nwmel_bank Project Status (01/26/2012 - 23:56:39) | | | |
|---|---|---|---|
| Project File: | mel.xise | Parser Errors: | No Errors |
| Module Name: | nwbark_bank | Implementation State: | Synthesized |
| Target Device: | xc4vfx12-10ff668 | • Errors: | No Errors |
| Product Version: | ISE 12.3 | • Warnings: | 49 Warnings (20 new) |
| Design Goal: | Balanced | • Routing Results: | |
| Design Strategy: | Xilinx Default (unlocked) | • Timing Constraints: | |
| Environment: | System Settings | • Final Timing Score: | |

| Device Utilization Summary (estimated values) | | | [-] |
|---|---|---|---|
| Logic Utilization | Used | Available | Utilization |
| Number of Slices | 965 | 5472 | 17% |
| Number of Slice Flip Flops | 1611 | 10944 | 14% |
| Number of 4 input LUTs | 1560 | 10944 | 14% |
| Number of bonded IOBs | 33 | 320 | 10% |
| Number of FIFO16/RAMB16s | 22 | 36 | 61% |
| Number of GCLKs | 1 | 32 | 3% |
| Number of DSP48s | 11 | 32 | 34% |

Figure 19. Device utilization summary of Bark filter bank

A comparative study of device utilization of Bark filter bank and Mel filter bank is given in Figure 20.In the comparison it is shown that Bark filter occupies less device compare to Mel filter bank.

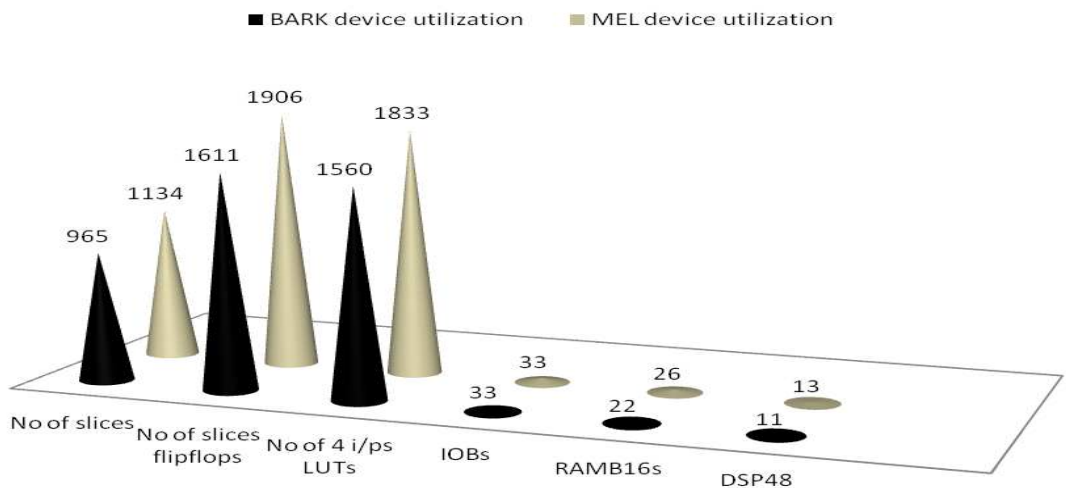

Figure 20 . Comparison of device utilization summary of Bark filter bank and Mel filter bank

## 3. CONCLUSIONS

This paper presents Mel filter bank and Bark filter bank which are portable among different EDA tools and technology independent. The whole designs are implemented in VHDL through Xilinx ISE11.1.Mel and Bark scale are frequency scales commonly found in psychoacoustics, i.e. they reflect how our ear detects pitch. They are approximately linear below 0.5 kHz and approximately logarithmic above that. Mel scale is reflecting our perception of pitch: equal subjective pitch increments produce equal increments in screen coordinate. Bark scale is reflecting our subjective loudness perception and energy integration. It is similar to Mel scale, but puts more emphasis on low frequencies. This filter bank has been integrated as part of a Speech Recognition System together with the other parts of the system such as end point detection, MFCC feature extraction. In this case the physical resources performance in order to have full implementation of the system in the same FPGA is more important than other criteria used. The designs are currently under final FPGA realization and will be reported in the future.

## ACKNOWLEDGEMENTS

The authors would like to thank the Electronics and Communication Engineering Department of Techno India, Salt Lake for the support provided High tech Lab facilities, Xilinx for generously donating the FPGAs and software used in this project, and Faculty members for help and advice.

## Authors

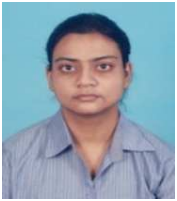
**Debalina Ghosh** is M.Tech perusing student in  Microelectronics and VLSI Design from Techno India, Salt Lake, India (2010-2012).She has done her B.Tech in Electronics And Communication Engineering from Academy Of Technology,Adisaptagram,India (2006-2010). Her present research area  is Embedded design and Digital Signal Processing

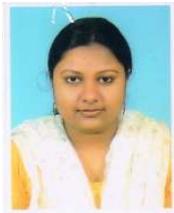
**Mrs. Depanwita Debnath ,**  presently  is designated as  an Assistant  Professor in the Electronics and Communication Department, Techno India, Saltlake , India. She has done her B.Tech in Electronics and Communication Engineering From College of Engineering and Management, Kolaghat, India (2003-2007) and M.Tech in Microelectronics and VLSI Design from IIT Kharagpur ( 2007-2009). Her present research area is Embedded design Digital Signal Processing.

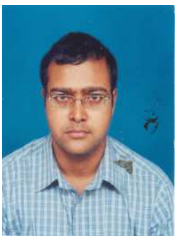
 **Mr Saikat Bose** presently is designated as an Assistant Professor in the Electronics and Communication Department,    Techno India, Saltlake , India. He has done his diploma in ECT from RKMS in 2000. AMITE in Electronics and Communication Engineering From New Delhi, India (2000-2004) and M.E in Embedded Systems from WBUT (2004-2006). His present research area is Embedded design Digital Signal Processing.